\documentclass[journal,comsoc]{IEEEtran}

\usepackage{amsmath}
\usepackage{amssymb}
\usepackage{amsfonts}
\usepackage{color}
\usepackage{graphicx}
\usepackage{algorithm}
\usepackage{algorithmic}
\usepackage{enumerate}
\usepackage{cite}
\usepackage{multirow}

\allowdisplaybreaks
\begin{document}
\title{Reliability-Based Windowed Decoding for Spatially-Coupled LDPC Codes}
\author{\IEEEauthorblockN{Peng Kang, Yixuan Xie, Lei Yang, and Jinhong Yuan}\vspace*{-10mm}
\vspace*{-1mm}\thanks{Peng Kang, Yixuan Xie, Lei Yang and Jinhong Yuan are with the School of Electrical Engineering and Telecommunications, the University of New South Wales, Sydney, Australia (email: peng.kang@unsw.edu.au; yixuan.xie@unsw.edu.au; lei.yang3@unsw.edu.au; j.yuan@unsw.edu.au, \textit{Corresponding author: Lei Yang.}).}}
%
\maketitle
\begin{abstract}
  In this letter, we propose a reliability-based windowed decoding scheme \color{black} for spatially-coupled (SC) low-density parity-check (LDPC) codes.
  To mitigate the error propagation along the sliding windowed decoder of the SC LDPC codes, a partial message reservation (PMR) method is proposed where only the reliable messages generated in the previous decoding window are reserved for the next decoding window.
  We also propose a partial syndrome check (PSC) stopping rule for each decoding window, in which only the complete VNs are checked.
  Simulation results show that our proposed scheme significantly improves the error floor performance compared to the sliding windowed decoder with the conventional weighted bit-flipping (WBF) algorithm.
\end{abstract}

\vspace*{-3mm}
\section{Introduction}
Spatially-coupled (SC) low-density parity-check (LDPC) codes \cite{Mitchell2015scldpc}, which are the convolutional counterparts of LDPC codes \cite{Gallager1962LDPC}, have drawn attention of many researchers.
An applicable way to decode an SC LDPC code is to use a sliding windowed decoder \cite{Iyengar2012window, Zhu2017Braided}.
Compared to the full block decoding (FBD) which decodes the entire codeword of an SC LDPC code with full flooding schedule \cite{Kschischang1998fsd}\cite{Xie2016EGSC}, the sliding windowed decoder shifts along the Tanner graph and focuses on decoding only a portion of a codeword at a time, which results in a lower decoding latency and memory requirement.
Since the windowed decoding architecture causes performance degradation compared to the FBD\color{black} \cite{Lentmaier2011Efficient}, most of the previous work, such as \cite{Ali2017iwd, Hassan2017nonWD, Lentmaier2011Efficient}, focused on improving the performance of the sliding windowed decoder with soft-decision decoding algorithms such as sum-product algorithm (SPA).
It is well-known that SPA leads to a high decoding complexity as soft information is passed along the edges in the Tanner graph \cite{Mohsenin2010routing}.
Therefore, the reliability-based decoding algorithms which only pass hard information along the edges in the Tanner graph, are investigated by many researchers to obtain a lower decoding complexity with acceptable performance degradation \cite{ryan2009channel, Nhan2015multiVote, Liu2005FGwbf}.
As shown in \cite{ryan2009channel}, the conventional weighted bit-flipping (WBF) algorithm assigns fixed weights obtained from received signals to the checksums.
The algorithm proposed in \cite{Nhan2015multiVote} allocates multiple voting levels to the unsatisfied checksums in order to improve the decoding performance.

\color{black}
In this letter, we investigate sliding windowed decoder for SC LDPC codes with the conventional WBF algorithm \cite{ryan2009channel}.
We observe that there is a significantly high error floor when the conventional WBF algorithm is used for windowed decoding of SC LDPC codes.
Since a sliding windowed decoder only covers a portion of the full Tanner graph, there exists variable nodes (VNs) that have neighbouring check nodes (CNs) outside the decoding window.
Thus, the messages sent out from these VNs may not be reliable.
These unreliable messages are propagated to the next window and deteriorate the error rate performance of the code.
Motivated by this observation, we propose a new approach to perform windowed decoding by only reserving the reliable messages between two adjacent windows.
In addition, we consider an improved stopping rule for the windowed decoding scheme.
We demonstrate that the proposed reliability-based windowed decoding (RBWD) scheme can significantly reduce the error floor of the SC LDPC codes constructed from protographs.
More importantly, the bit error rate (BER) performance of the RBWD scheme approaches that of FBD within $0.1$ dB, which is highly desirable for applications with a low decoding complexity requirement.
\vspace*{-3mm}
\section{Windowed Decoding and WBF Algorithm}
\subsection{Sliding Windowed Decoder for SC LDPC Codes}
\vspace*{-1mm}
Let $\mathbf{B}$ be a base matrix of size $r\times c$ for a $(J,cJ)$-regular protograph LDPC code, where $r$ and $c$ are the number of rows and columns in $\mathbf{B}$, respectively.
The base matrix of an SC LDPC code can be generated from $\mathbf{B}$ as
\begin{align}\label{equ:SCLDPC}
\small
\setlength{\arraycolsep}{1pt}
\renewcommand{\arraystretch}{0.01}
\mathcal{B}_L = \left[
{\overbrace {\begin{array}{*{10}{cccc}}
{{{\bf{B}}_0}}&{}&{}&{}\\[-1mm]
 \vdots &{{{\bf{B}}_0}}&{}&{}\\[-2mm]
{{{\bf{B}}_{{m_s} - 1}}}& \vdots & \ddots &{{{\bf{B}}_0}}\\[-2mm]
{{{\bf{B}}_{{m_s}}}}&{{{\bf{B}}_{{m_s} - 1}}}& \ddots & \vdots \\[-2mm]
{}&{{{\bf{B}}_{{m_s}}}}& \ddots &{{{\bf{B}}_{{m_s} - 1}}}\\[2mm]
{}&{}&{}&{{{\bf{B}}_{{m_s}}}}\\[-2mm]
\end{array}}^L} \right],\\[-4mm]
\begin{array}{*{10}{c}}
\hspace*{5mm}\,\,\,1&{}&2&{}&\hdots&L&{}&{}\vspace*{-5mm}
\end{array}\notag\vspace*{-3mm}
\end{align}
%
where $m_s$ is called syndrome former memory and $L$ is known as termination length.
Each $\mathbf{B}_j$ of size $r\times c$ is a descendent matrix of $\mathbf{B}$ such that $\sum_{j=0}^{m_s}\mathbf{B}_j = \mathbf{B}$, where the set of matrices $\mathbf{B}_j$ is obtained by performing edge spreading \cite{Mitchell2015scldpc}.
In this letter, we consider full edge spreading \cite{Mitchell2015scldpc} for the construction of SC LDPC codes, i.e., $\mathbf{B}_0 = \mathbf{B}_{1} = \cdots = {\mathbf{B}}_{m_s}=[1\;1\; \cdots \;1]_{1 \times c}$, where $m_{s} = J - 1$.
After edge spreading and graph expansion operation with lifting size $M$, a full parity-check matrix of an SC LDPC code can be obtained.

In \cite{Iyengar2012window}, a sliding windowed decoder was proposed.
Instead of performing FBD over the whole base matrix $\mathcal{B}_L$, the sliding windowed decoder uses a window of size $W$ covering $W \cdot Mr$ CNs and $W \cdot Mc$ VNs.
The decoding window slides from time index $t=1$ to time index $t=L$ which associates with different window positions in $\mathcal{B}_L$.
In a decoding window, an iterative message-passing decoding algorithm is performed between all VNs and CNs.
The decoding process stops if a valid codeword is found or a predetermined maximum number of iterations is reached.
Then the decoding window shifts by $Mr$ CNs vertically and $Mc$ VNs horizontally where the $Mc$ VNs shifted out of the decoding window are called target symbols.
\vspace*{-4mm}
\subsection{The Conventional WBF Algorithm}
\vspace*{-1mm}
Denoted by $\mathbf{H}$ an $m \times n$ parity-check matrix.
For $0 \leq j \leq m-1$ and $0 \leq i \leq n-1$, let $\mathcal{M}(j)$ and $\mathcal{N}(i)$ be the sets of indices of all the nonzero elements in the $j$-th row and $i$-th column of $\mathbf{H}$, respectively.
Define $\mathbf{r} = ({r_0},{r_1}, \ldots {r_{n - 1}})$ as the soft-decision received sequence at the channel output.
Let $\mathbf{v} = ({v_0},{v_1}, \ldots {v_{n - 1}})$ be the hard decision sequence obtained from $\mathbf{r}$ as follow: $v_{i}=0$ for $r_{i}>0$ and $v_{i}=1$ for $r_{i} \leq 0$.
Denoted by $\mathbf{s}^{(l)} = (s_0^{(l)},s_1^{(l)}, \ldots s_{m - 1}^{(l)})$ the syndrome vector computed for the flipping metric at the $l$-th iteration.
The conventional WBF algorithm computes the flipping metric of each VN at the $l$-th iteration according to \cite{ryan2009channel}
\begin{equation}\label{reliabilityFunc}
{E_i^{(l)}} = \sum\nolimits_{j \in \mathcal{N}(i)} {(2{s_j^{(l)}} - 1) \cdot {w_j}},\vspace{-1mm}
\end{equation}
where $w_j$ is a weighted factor given by ${w_j} = \mathop {\min }\limits_{i \in \mathcal{M}(j)} \left| {{r_i}} \right|.$
Then the candidate bit(s) to be flipped can be determined by
\begin{equation}\label{flipRule}
\mathbb{F} = \{ i|i = \mathop {\arg \max E_i^{(l)}}\limits_{0 \le i \le n - 1} \}.\vspace{-1mm}
\end{equation}
The process repeats until all the parity-check equations are satisfied or a preset maximum number of iterations is reached.
As shown in \cite{ryan2009channel}, the conventional WBF algorithm may flip multiple bits selected from $\mathbb{F}$ in one iteration.
Although the multi-bit flipping (MBF) rule leads to a fast convergence speed, carefully designed loop removal mechanisms are required to avoid the decoding process to be trapped in an infinite loop due to its greediness \cite{Liu2005FGwbf}.
An alternative flipping rule for the conventional WBF algorithm is to randomly flip one bit in $\mathbb{F}$ at each iteration \cite{ryan2009channel}, which is also called the single-bit flipping WBF (SBF-WBF) decoding algorithm.
\vspace*{-2mm}
\section{Reliability-based Windowed Decoding for SC LDPC Codes}
By employing the sliding windowed decoder with the conventional WBF algorithm for SC LDPC codes, we observe a considerable performance loss caused by error floor.
In this section, we propose a partial message reservation (PMR) method and a partial syndrome check (PSC) stopping rule for the windowed decoder, to solve this problem.
\vspace*{-3mm}
\subsection{The PMR Method}
\vspace*{-1mm}
Due to the structure of the sliding windowed decoder, we observe that some of the VNs in the decoding window have neighboring CNs outside the window.
We call these VNs as $\emph{incomplete}$ VNs and the others as $\emph{complete}$ VNs for this decoding window.
It was shown in \cite{ryan2009channel} that the performance of the conventional WBF algorithm highly relies on a large column weight of the given parity-check matrix for an LDPC code.
However, in the construction defined by Eq. (\ref{equ:SCLDPC}), the incomplete VNs have a lower column weight than that of complete VNs.
Therefore, the messages passed along the edges connected to the incomplete VNs are less reliable than that associated with the complete VNs.

It is well-known that the good performance of SC LDPC codes with windowed decoding comes from reliable messages passed from one window to the next.
To avoid the error propagation of unreliable messages from the incomplete VNs, we propose a PMR method for the sliding windowed decoder.
Let $\mathbb{V}_{C}$ and $\mathbb{V}_{I}$ represent the sets of indices for complete VNs and incomplete VNs in a decoding window, respectively.
Define $\mathbf{z}_{t} = (z_{t,0}, z_{t,1}, \ldots z_{t,n'-1})$ as the decoded codeword for the current window at time index $t$, where $n' = W \cdot Mc$.
The outgoing message from the $k$-th VN in the current decoding window to the next window can be given by
\begin{equation}\label{msgReserve}
\hspace*{-3mm}{u_{k}} = \left\{\hspace*{-3mm}\begin{array}{c}
\begin{array}{*{20}{c}}
{{z_{t,{k}}},}&{{k} \in \mathbb{V}_{C}}
\end{array}\\
\begin{array}{*{20}{c}}
{{v_{k}},}&{\,\,\,{k} \in \mathbb{V}_{I}}
\end{array}
\end{array} \right.,\vspace{-1mm}
\end{equation}
where $0 \leq k \leq n'-1$.
This means that only the messages from complete VNs in the $t$-th window are reserved for the $t+1$-th window.
Note that the window size is chosen to ensure that the number of complete VNs in a decoding window is larger than that of incomplete VNs, so that more reliable messages can be reserved and propagated to the next window.
We will show in Fig. 3 that this PMR method can significantly improve the error floor performance of the proposed RBWD scheme.
\vspace*{-2mm}
\subsection{The PSC Stopping Rule}
\vspace*{-1mm}
When decoding an LDPC code with parity-check matrix of size $m\times n$, all the $m$ parity-check equations need to be satisfied to get a valid codeword.
An efficient stopping rule based on soft bit error indicators was introduced in \cite{Hassan2017nonWD} for sliding windowed decoder.
However, this method can not be directly applied to a RBWD scheme since only hard information is passed along the edges in the Tanner graph.

By making use of the reliable messages, a PSC stopping rule is applied to the windowed decoding scheme.
In particular, our stopping rule only focuses on the parity-check equations of complete VNs in a decoding window.
To be specific, define $W'$ as the number of parity-check equations in one decoding window considered by the PSC stopping rule.
The first $W'=(W-m_{s}) \cdot Mr$ parity-check rows are checked in each decoding window.
Once these parity-check equations are satisfied or the preset maximum number of iterations is reached, the decoding window slides to the next position.
Note that a PSC stopping rule is also proposed in \cite{Ali2017iwd}.
However, it only checks a fixed number of syndromes that belong to the target symbols.
In our proposed PSC stopping rule, all reliable VNs are considered. When $W > m_{s}+1$, the number of complete VNs in one decoding window is larger than that of target symbols, which leads to a more strict stopping rule and ensures the messages from complete VNs to be more reliable.

\begin{figure}
  \centering
  \includegraphics[width=1.5in]{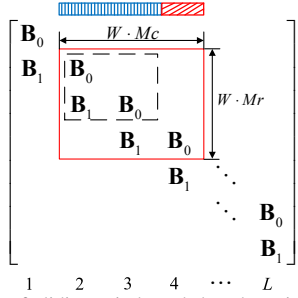}
  \vspace*{-3.5mm}
  \caption{An example of sliding windowed decoder with window size $W=3$ at time index $t=2$ (solid region). The parity-check equations considered by the PSC stopping rule are shown in dashed region. The complete VNs are shown in blue (vertically hatched) and the incomplete VNs are shown in red (hatched) above the parity-check matrix.}\vspace*{-7mm}
  \label{WDex}
\end{figure}

An example of the proposed sliding windowed decoder with $W=3$ and $m_s=1$ at time index $t=2$ is illustrated in Fig. \ref{WDex}.
The first $2Mc$ VNs are complete VNs and the last $Mc$ VNs are incomplete VNs.
Note that only the first $2Mr$ CNs are considered for the parity-check equations since these CNs connect to the complete VNs.
The last updated messages of the first $2Mc$ complete VNs are reserved for the decoding process at time index $t=3$.
\vspace*{-2mm}
\subsection{The Proposed RBWD Scheme}
\begin{figure}[h!]
  \centering
  \vspace*{-5mm}
  \includegraphics[width=2.5in]{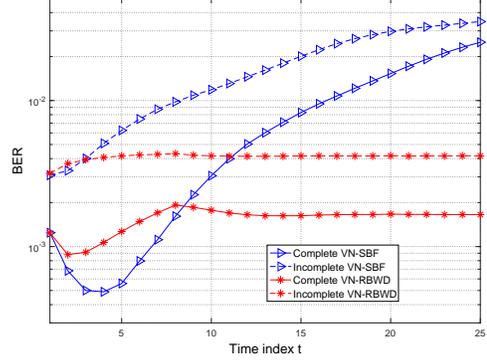}
  \vspace*{-3mm}
  \caption{Error performance of complete \& incomplete VNs for the $(7,49)$ SC LDPC code at different time $t$, $E_{b}/N_{0}$ = $6$ dB.}\vspace*{-4mm}
  \label{errPos}
\end{figure}

\vspace*{-1mm}
Denoted by $\mathbf{\hat{H}}$ an $m' \times n'$ parity-check matrix for one decoding window, where $m' = W \cdot Mr$ and $n' = W \cdot Mc$.
Let $\mathbf{s'}^{(l)} = ({s'}_0^{(l)},{s'}_1^{(l)}, \ldots {s'}_{W' - 1}^{(l)})$ be the syndrome vector computed by the PSC stopping rule at the $l$-th iteration.
Assume that vector $\mathbf{y}_{t}^{(l)} = (y_{t,0}^{(l)},y_{t,1}^{(l)}, \ldots y_{t,n' - 1}^{(l)})$ is the decoded codeword of the $l$-th iteration at time index $t$.
Define $\mathcal{M}'(j')$ and $\mathcal{N}'(i')$ as the sets of indices of all the nonzero elements in the $j'$-th row and $i'$-th column of $\mathbf{\hat{H}}$, respectively.
Set the maximum number of decoding iterations as $I_{max}$.
By combining the PMR and PSC with the SBF-WBF algorithm, the proposed RBWD scheme is summarized in \textbf{Algorithm \ref{alg:rbwd}}.
\vspace*{-2mm}
\begin{algorithm}
 \caption{The proposed RBWD scheme}\label{alg:rbwd}
 \begin{algorithmic}[1]
 \renewcommand{\algorithmicrequire}{\textbf{Inputs:}}
 \renewcommand{\algorithmicensure}{\textbf{Output:}}
 \REQUIRE $\mathbf{\hat{H}}, L, W, M, I_{max}$
 \STATE \textbf{Initialize:} $l=0$ and $t=1$\vspace*{-1mm}
 \WHILE {$t \leq L$}\vspace*{-1mm}
 \IF {$t=1$}
 \STATE {set $\mathbf{y}_{1}^{(0)} = \mathbf{v}$}\vspace*{-1mm}
 \ELSE\vspace*{-1mm}
 \STATE {set $y_{t,i'}^{(0)} = \left.\Big\{{\begin{array}{rcl}
   \vspace*{-1mm} {{v_{i'}},}&\,\hspace*{2mm}n' - 1 - Mc \leq i' < n' - 1\\
   \vspace*{1mm}{{u_{i'}},}&\hspace*{-4mm}0 \leq i' < n' - 1 - Mc
   \end{array}} \right.$}\vspace*{-1mm}
 \ENDIF
 \WHILE {$l \leq I_{max}$}
 \FOR {$j'=0:(m'-1)$}\vspace*{-1mm}
 \STATE {${w_j} = \mathop {\min }\limits_{i \in \mathcal{M}(j)} \left| {{r_i}} \right|$}\vspace*{-1mm}
 \ENDFOR
 \STATE Update $l=l+1$
 \STATE {Compute ${s^{(l)}}$ by $\mathbf{y}_{t}^{(l)}\hat{H}^{T}$}
 \STATE {Determine $s'^{(l)}=(s_0^{(l)},s_1^{(l)}, \ldots s_{W' - 1}^{(l)})$}
 \IF {${s'^{(l)}} = \mathbf{0}$ or $l = I_{max}$}
 \STATE {output $\mathbf{z}_{t}=\mathbf{y}_{t}^{(l)}$ and \textbf{break}}
 \ENDIF
 \FOR {$i'=0:(n'-1)$}
 \STATE Estimate ${E_i^{(l)}}$ as in (2)
 \ENDFOR
 \STATE Update $\mathbb{F}$ as in (3)
 \STATE Flip $y_{t,i'}^{(l)}$ randomly, where $i' \in \mathbb{F}$
 \ENDWHILE
 \STATE Perform PMR as in (4), set $t=t+1$ and $l = 0$
 \ENDWHILE
 \end{algorithmic}
 \end{algorithm}
\vspace*{-3mm}
Note that the performance gain of the proposed RBWD scheme originates from the discarding of unreliable messages from previous decoding window to perform message-passing decoding in the current window.
To demonstrate this, we evaluate the BER performance of complete and incomplete VNs for various window positions.
Fig. \ref{errPos} depicts the BER for both complete and incomplete VNs of an SC LDPC code constructed from a $(7,49)$-regular protograph LDPC code with $m_{s} = 6$, $L=56$ at different window positions.
The decoding window size $W$ is set to $14$.
It can be seen that for both the SBF-WBF algorithm and the proposed RBWD decoding scheme, incomplete VNs always have a higher BER than complete VNs.
For instance, the BER of incomplete VNs by using RBWD scheme is nearly two times as that of the complete VNs for all window positions.
The difference of BER for those two types of VNs can be even larger for the SBF-WBF algorithm.
This indicates that the messages from incomplete VNs are less reliable than that from complete VNs.

It can also be seen that the BER for both complete and incomplete VNs of the SBF-WBF algorithm increases with the sliding of the decoding window.
For example, the BER of complete and incomplete VNs for the SBF-WBF algorithm increases more than ten times from the first decoding window to the $25$-th decoding window for the simulated SC LDPC code.
On the other hand, the BER of both complete and incomplete VNs for the proposed RBWD scheme remains almost the same for all decoding window positions.
This means that by only reserving the messages from the complete VNs in the RBWD scheme, we prevent the ``contamination" of the reliable messages from the unreliable messages.
As a result, the BER performance of the proposed decoding scheme is substantially improved.
\vspace*{-2mm}
\section{Numerical Results}\label{result}
%
In this section, we investigate the error rate performance and the decoding complexity of the proposed RBWD scheme.
Binary phase shift keying (BPSK) modulation and additive Gaussian noise channels are considered in all simulations.
The maximum number of iterations is 200 for all windowed decoding schemes and it is 2000 for FBD.
\vspace*{-4mm}
\subsection{Error Rate Performance}
\begin{figure}
  \centering
  \includegraphics[width=2.75in]{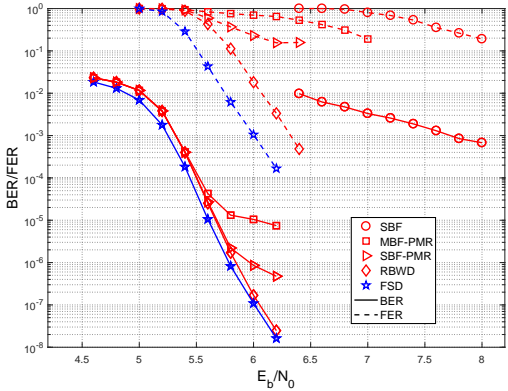}
  \vspace*{-3.5mm}
  \caption{BER/FER performance of the length-38024 $(7,49)$ SC LDPC code. The window size is $W=14$ for windowed decoding.}
  \vspace*{-2mm}
  \label{WD14}
\end{figure}

\vspace*{-1mm}
An SC LDPC code is constructed from a $(7,49)$-regular protograph LDPC code with full edge-spreading, i.e., $\mathbf{B}_0 = \mathbf{B}_{1} = \cdots = {\mathbf{B}}_{6}=[1\;1\; \cdots \;1]_{1 \times 7}$.
We set the termination length $L=56$, the resultant base matrix $\mathcal{B}_L$ is expanded with lifting size $M=97$.
As a result, we obtain a length-38024 $(7,49)$ SC LDPC code with large VN degrees.
The BER and frame error rate (FER) of the length-38024 $(7,49)$ SC LDPC code decoded by various decoding schemes are shown in Fig. \ref{WD14}.
Here MBF-PMR and SBF-PMR represent the RBWD scheme without applying the proposed PSC stopping rule.
The BER and FER of the FBD and the sliding windowed decoder based on the SBF-WBF algorithm are also shown in the figure for comparison.
We see that the proposed PMR method dramatically improves the error rate performance.
Moreover, the proposed stopping rule further reduces the error floor and achieves the BER performance within $0.1$ dB from that of the FBD.

\begin{figure}
  \centering
  \includegraphics[width=2.75in]{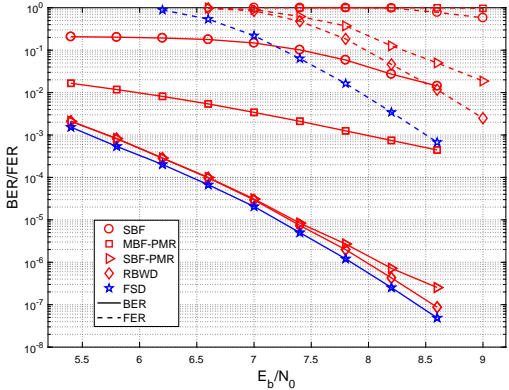}
  \vspace*{-3.5mm}
  \caption{BER/FER performance of the length-38016 $(3,6)$ SC LDPC code. The window size is $W=6$ for windowed decoding.}
  \vspace*{-6mm}
  \label{WD6}
\end{figure}

Note that the proposed RBWD scheme also works for SC LDPC codes with small VN degrees.
To clarify the generality, we constructed an SC LDPC code from a $(3,6)$-regular protograph LDPC code.
After applying the edge spreading matrices $\mathbf{B}_0 = \mathbf{B}_{1} = {\mathbf{B}}_{2}=[1\;1]_{1 \times 2}$ and set the termination length $L=108$, a length-38016 $(3,6)$ SC LDPC code can be obtained by graph expansion with lifting size $M=176$.
As shown in Fig. \ref{WD6}, the proposed RBWD scheme works for SC LDPC codes with small VN degrees in the sense that the BER performance of the RBWD scheme can approach that of the FBD.\vspace*{-1.5mm}
\vspace*{-1mm}
\vspace*{-1mm}
\subsection{Complexity Comparison}
\vspace*{-1mm}
In this section, we compare the complexity of the proposed RBWD scheme with that of the MBF-PMR and the SBF-PMR schemes.
Note that we only consider the decoding schemes based on the conventional WBF algorithm since it only exchanges one bit information between CNs and VNs, which has a lower decoding complexity than SPA.
In addition, we define $I_{avg}$ as the average number of updates processed by a VN in one decoded codeword, which can be given by
\vspace{-1mm}
\begin{equation}
\setcounter{equation}{6}
{I_{avg}} = \left(\sum\nolimits_{t = 1}^L {{I_t}}\right)/L,\vspace{-1mm}
\end{equation}
where $I_{t}$ is defined as the total number of updates processed by a VN at the $t$-th window during the decoding process.
The comparison of $I_{avg}$ for the length-38024 $(7,49)$ SC LDPC code decoded by MBF-PMR, SBF-PMR, and the proposed RBWD scheme is shown in Table \ref{Iavg}.
Note that for a fair comparison we fix $W=14$ for all windowed decoding schemes, i.e., each decoding window covers 9506 bits in order to keep the same decoding latency.
It can be seen that for SNR from $5.6$ dB to $6$ dB, our proposed RBWD scheme requires about half number of updates compared to that of MBF-PMR and SBF-PMR.
\begin{table}[t]
\centering
\caption{Average iteration comparison of the length-38024 $(7,49)$ SC LDPC code decoded by various decoding schemes.\vspace*{-2mm}}\label{Iavg}
\begin{tabular}{|c|c|c|c|c|c|c|c|}
\hline
\multicolumn{2}{|c|}{$E_{b}/N_{0}$ (dB)}                                                          & $5.2$ & $5.4$   & $5.6$ & $5.8$ & $6$  & $6.2$\\ \hline\hline
\multirow{3}{*}{$I_{avg}$}                    & MBF-PMR                                           & $132$ & $59$             & $29$  & $21$  & $17$ & 13  \\ \cline{2-8}
                                              & SBF-PMR                                           & $132$ & $57$             & $29$  & $22$  & $18$ & $15$  \\ \cline{2-8}
                                              & RBWD                         & \multicolumn{1}{c|}{$129$} & \multicolumn{1}{c|}{$48$} & $17$  & $11$  & $8$  & $7$ \\ \hline
\end{tabular}\vspace*{-6mm}
\end{table}
\vspace*{-3mm}
\section{Conclusion}
In this letter, we proposed a RBWD scheme for SC LDPC codes.
The proposed scheme propagates the reliable messages from complete VNs between two consecutive decoding windows, which substantially improves the error rate performance.
The proposed stopping rule in the RBWD scheme further reduces the error floor by operating on the parity-check equations that only involve complete VNs.
Numerical results showed that the proposed RBWD scheme can approach the BER performance of the FBD within $0.1$ dB.
\bibliographystyle{ieeetr}

\end{document}